# Substrate-controlled epitaxial-like growth of photochromic yttrium oxyhydride thin films

M. Zubkins[1*], E. Strods[1], E. Letko[1], H. Mändar[2], D. Moldarev[3], K. Mizohata[3], K. Sarakinos[3,4], V. Vibornijs[1], K. Kundzins[1], J. Purans[1]

[1]Institute of Solid State Physics, University of Latvia, Kengaraga 8, LV-1063, Riga, Latvia

[2]Institute of Physics, University of Tartu, W. Ostwald Str. 1, 50411, Tartu, Estonia

[3]Department of Physics, University of Helsinki, P.O. Box 43, FI-00014 Helsinki, Finland

[4]KTH Royal Institute of Technology, Department of Physics, Roslagstullsbacken 21, 11421 Stockholm, Sweden

*Corresponding author: martins.zubkins@cfi.lu.lv

**Abstract**

Yttrium oxyhydride (YHO) thin films are synthesised by reactive magnetron sputtering of $\beta$-$YH_x$ precursor films onto single-crystalline yttria-stabilised zirconia (YSZ) (111) and (100) substrates at 200 °C, followed by post-deposition oxidation in air. For comparison, reference films are also prepared on amorphous $SiO_2$ and soda-lime glass substrates at both 200 °C and room temperature (RT). The structural (X-ray diffraction and scanning electron microscopy), compositional (elastic recoil detection analysis), and optical properties, including photochromic behaviour, of the films are systematically investigated. The results reveal a strong influence of substrate orientation on the film microstructure and crystallographic texture, leading to epitaxial-like growth with a mosaic spread of ~3–4°. YHO deposited on YSZ(100) exhibits larger crystallite sizes (10–20 nm) than YHO grown on YSZ(111) (~8 nm) and exhibits residual in-plane compressive strain while remaining largely relaxed with respect to the substrate. Compositional analysis revealed that the O/Y and O/H atomic ratios ranged from 0.71 to 0.82 and from 0.33 to 0.44, respectively, for films deposited under optimised conditions. These films exhibit photochromic behaviour under solar simulator illumination, with relative transmittance changes ranging from 7 to 22 % after 10 min of illumination. Despite the pronounced structural differences, the epitaxial-like films show no systematic improvement in photochromic response, indicating that the effect is governed primarily by composition. The study shows that the bleaching time constant increases exponentially with increasing contrast reached during illumination, rather than representing an intrinsic material constant of YHO.

## Introduction

Yttrium oxyhydride (YHO) is an inorganic photochromic compound exhibiting reversible light-induced modulation of optical and electrical properties under ambient conditions [1]. In thin-film form, YHO combines high optical transparency in the clear state, with visible-light transmittance values typically ranging from 70 to 80 %, together with colour-neutral photo-darkening and spontaneous bleaching in the dark. The photochromic properties strongly depend on the composition [2], film thickness [3], microstructure, including local atomic arrangement and defects [4,5], as well as illumination conditions [6]. Since the composition of this compound can vary over a broad range, the abbreviation YHO is used hereafter instead of specifying a fixed stoichiometry unless stated otherwise. The optical band gap of transparent photochromic YHO films is generally reported in the range of 2.6–3.1 eV [7], while the photochromic response extends from the visible to the mid-infrared spectral region [8]. These properties make YHO promising for various optical and optoelectronic applications, including passive light-modulating coatings, optical switches, optical memory concepts, and photoresponsive electronic devices.

Structurally, YHO belongs to the family of mixed-anion materials [9], in which oxide ($O^{2-}$) and hydride ($H^{-}$) anions are incorporated into a face-centred cubic lattice [10]. Experimental and theoretical studies indicate that oxygen and hydrogen predominantly occupy tetrahedral interstitial sites, while excess hydrogen may additionally occupy octahedral positions [10,11]. The exact anion arrangement, local ordering, and defect structure remain subjects of active discussion because they are closely related to the origin of the photochromic effect. Several mechanisms have been proposed to explain the photochromic effect, including the formation of metallic domains [12], anion-vacancy-related electronic transitions [4], and hydrogen redistribution within the film [13].

YHO thin films are typically synthesised using a two-step process involving reactive magnetron sputtering of yttrium hydride ($\beta$-$YH_x$, $x$ = 1.8–2.3 [14,15]) followed by post-deposition oxidation in air or controlled oxygen-containing atmospheres [16,17]. This approach is compatible with scalable thin-film deposition technologies [18]. Most previous studies have focused on polycrystalline films deposited on amorphous substrates such as glass, whereas investigations of epitaxial YHO films remain very limited.

Epitaxial growth enables the formation of thin films with well-defined crystallographic orientation and reduced structural disorder compared to polycrystalline materials. Epitaxial films provide a model system for investigating intrinsic material properties without the dominant influence of randomly oriented grains and high-density grain boundaries. In mixed-anion materials such as YHO, epitaxial growth is particularly relevant because it may influence anion ordering, defect formation, hydrogen mobility, and strain-induced structural distortions, all of which are closely related to the photochromic response.

To date, only Komatsu et al. [19] have reported the epitaxial growth of YHO films deposited on single-crystalline yttria-stabilised zirconia (YSZ) (111) and $CaF_2$(111) substrates at elevated temperature, demonstrating a photoinduced insulator-to-metal transition under UV laser illumination. However, detailed analysis of the epitaxial quality, anisotropy, and photochromic properties of epitaxial YHO films has remained largely unexplored. Furthermore, deposition on differently oriented crystalline substrates, such as YSZ(100), has not yet been investigated, and the growth of YHO thin films at elevated temperatures has received limited attention.

The primary objective of this study is to investigate the structural properties and epitaxial quality of YHO thin films deposited on differently oriented single-crystalline substrates, including their lattice orientation relationships, mosaicity, anisotropy, and microstructure. The optical and photochromic properties are discussed in the context of these structural characteristics together with the chemical composition to evaluate whether epitaxial ordering influences the photochromic response.

To this end, YHO thin films were deposited on single-crystalline YSZ(111) and YSZ(100) substrates by reactive magnetron sputtering at an elevated substrate temperature of 200 °C, followed by post-deposition oxidation in air. The structural properties and epitaxial quality of the films were investigated using high-resolution X-ray diffraction (HR-XRD) and scanning electron microscopy (SEM). In addition, the relationship between microstructure, composition determined by time-of-flight energy elastic recoil detection analysis (ToF-E ERDA), and photochromic response was analysed and compared with that of textured polycrystalline films deposited on amorphous $SiO_2$ and soda-lime glass substrates at both 200 °C and room temperature (RT).

The results demonstrate epitaxial-like growth on both YSZ orientations and reveal pronounced substrate-dependent differences in crystallographic quality, lattice anisotropy, and microstructure. Despite these structural differences, no systematic enhancement of the photochromic response is observed for the epitaxial-like films, indicating that photochromic contrast is governed primarily by film composition. Furthermore, the bleaching time constant increases strongly with the photochromic contrast reached during illumination and therefore should not be regarded as an intrinsic material constant of YHO.

**Experimental details**

YHO thin films were synthesised on YSZ substrates with (100) and (111) orientations. A two-step synthesis route was employed, consisting of (i) deposition of $\beta$-$YH_x$ and (ii) subsequent post-oxidation in air. The $\beta$-$YH_x$ precursor films were deposited by reactive pulsed direct (80 kHz, 2.5 μs reverse time) current magnetron sputtering at a substrate temperature of 200 °C. For comparison, reference films were also deposited at 200 °C and RT (without intentional substrate heating) on $SiO_2$ and soda-lime glass substrates.

The deposition conditions for the films grown on YSZ substrates were systematically optimised with three objectives: (i) to obtain transparent YHO films exhibiting the highest possible photochromic contrast, (ii) promote out-of-plane lattice matching of YHO with the YSZ substrates, and (iii) minimise the well-known delamination of YHO films from the substrate [17,20]. In contrast, the reference films were optimised primarily for photochromic performance and stability.

The deposition pressure of the $\beta$-$YH_x$ precursor is a critical parameter governing the oxidation behaviour and, consequently, the optical and photochromic properties of YHO films [21]. Films deposited below the critical pressure ($P_c$) form dense precursor structures that hinder oxygen incorporation during post-deposition oxidation, thereby preventing complete conversion into the transparent photochromic YHO phase. Conversely, films deposited at pressures significantly above $P_c$ exhibit enhanced oxygen incorporation and higher optical transmittance, although at the expense of photochromic performance. Previous studies have shown that the maximum photochromic contrast is achieved for films deposited at pressures slightly above $P_c$ [7,22]. Furthermore, improved photochromic properties can be achieved by using the lowest $H_2$/Ar ratio that ensures single-phase $\beta$-$YH_x$ growth without the formation of a mixed metallic-like $\alpha$-Y/ $\beta$-$YH_x$ phase [23,24].

The growth of $\beta$-$YH_x$ precursor films was found to depend strongly on both the substrate type and the deposition temperature. Consequently, the working pressure was varied between 0.65 and 1.80 Pa to position each deposition condition close to its corresponding $P_c$. Increasing the substrate temperature to 200 °C shifted $P_c$ towards higher values compared with the room-temperature reference samples, for which the optimal deposition pressure was 0.65 Pa. At 200 °C, $P$c was approximately 0.95 Pa for films deposited on YSZ(100) and $SiO_2$, whereas a substantially higher pressure of 1.80 Pa was required for growth on YSZ(111).

Deposition at pressures exceeding these optimum values resulted in progressively stronger film delamination, which developed during post-oxidation and became more pronounced upon prolonged exposure to ambient conditions. Finally, to further improve the crystallographic quality of the films deposited at 200 °C, the discharge power was reduced to 100 W to decrease the deposition flux and minimise ion-induced defect formation during film growth.

Following deposition at 200 °C, the yttrium hydride films were maintained in the vacuum chamber under an argon atmosphere until the substrate temperature decreased to 80–85 °C, since the YHO phase is not stable at temperatures above 100 °C [2,25]. The chamber was then vented to air, initiating oxidation and leading to the formation of the YHO phase. The representative samples investigated in this study, including the optimised ones, together with their deposition parameters, are summarised in Table 1.

**Table 1.** Deposition parameters of YHO thin films prepared on single-crystalline YSZ (111) and YSZ (100), as well as amorphous $SiO_2$ and soda-lime glass substrates. The table summarises the sample identifiers, substrate type, substrate temperature, working pressure, $H_2$/Ar gas flow ratio, discharge power, and resulting film thickness.

| Sample | Substrate | Temp. (°C) | Pressure (Pa) | $H_2$/Ar | Power (W) | Thickness (nm) |
|---|---|---|---|---|---|---|
| S-YSZ111* | YSZ(111) | 200 | 1.80 | 0.05 | 100 | 350 |
| S-YSZ100-1* | YSZ(100) | 200 | 0.95 | 0.05 | 100 | 325 |
| S-YSZ100-2 | YSZ(100) | 200 | 1.65 | 0.02 | 200 | 620 |
| S-SiO-200C* | $SiO_2$ | 200 | 0.95 | 0.05 | 100 | 390 |
| S-SiO-RT | $SiO_2$ | RT | 0.70 | 0.13 | 200 | 465 |
| S-SL-RT* | Soda-lime | RT | 0.65 | 0.01 | 200 | 550 |

* Optimised samples

The film thickness was determined using a WOOLLAM RC2 spectroscopic ellipsometer over the spectral range of 0.73–2.25 eV, where the films are highly transparent. The optical response within this energy range was modelled by the WOOLLAM CompleteEASE software [26] using a Cauchy dispersion relation $n(\lambda) = A + B\lambda^{-2}$. The optical constants of the YSZ substrates were determined beforehand, whereas those of the $SiO_2$ and soda-lime glass substrates were taken from the software library.

All samples were initially characterised using a Rigaku MiniFlex600 X-ray powder diffractometer operating in Bragg–Brentano $\theta$–$2\theta$ geometry with a 600 W Cu anode X-ray tube (Cu Kα radiation, $\lambda$ = 1.5406 Å). These measurements provided phase identification and qualitative texture information and were used to screen the samples prior to detailed high-resolution X-ray diffraction (HR-XRD) analysis of selected films. The degree of preferred orientation in the films deposited on amorphous substrates was quantified using the texture coefficient (*TC*), calculated according to:

$$TC_{hkl} = \frac{I_{hkl}/I^*_{hkl}}{\frac{1}{n}\Sigma(I_{hkl}/I^*_{hkl})}, \tag{1}$$

where $I_{hkl}$ is the measured intensity of the diffraction peak corresponding to the (*hkl*) plane, $I^*_{hkl}$ is the corresponding reference intensity of cubic $YH_2$ (ICDD card No. 04-002-6939), and $n$ is the number of reflections included in the calculation. In the present study, the *TC* values were calculated using the 111, 200, 220, and 311 peaks.

HR-XRD was subsequently employed to analyse the lattice orientation and mosaicity (both out-of-plane and in-plane) of the selected films on both orientations of the YSZ substrates using a Rigaku SmartLab™ diffractometer. The measurements were performed in parallel beam geometry using monochromatized (Ge(220) two-bounce monochromator) Cu $K\alpha_1$ X-ray radiation ($\lambda$ = 0.1540598 nm) from rotating anode operated at 45 kV and 180 mA.

The surface morphology and cross-sections of the films were examined using a Helios 5 UX dual-beam scanning electron microscope (SEM; Thermo Fisher Scientific). The images were acquired at an acceleration voltage of 5 kV using an Elstar in-lens SE detector.

Chemical composition of the films was determined by time-of-flight energy elastic recoil detection analysis (ToF-E ERDA). The measurements were conducted at the 5 MV tandem accelerator of the Helsinki Accelerator Laboratory (University of Helsinki) using a 50 MeV $^{127}I^{9+}$ ion beam. During analysis, the samples were tilted by 20° relative to the incident beam, while the time-of-flight telescope and energy detectors were positioned at a scattering angle of 40°. Signals from both detectors were recorded in coincidence mode [27], and the acquired spectra were evaluated using the Potku software [28].

Optical transmittance spectra were measured using a UV–Vis–NIR spectrophotometer (Agilent Cary 7000) over the wavelength range of 300–2500 nm. The photochromic response was evaluated under ambient conditions using an Oscilla LED solar simulator (100 mW $cm^{-2}$) as the illumination source. Optical probing was performed with a broadband white light source covering the 360–900 nm spectral range. To minimise any unintended photo-darkening induced by the probe light, a 480 nm long-pass filter was introduced in the optical path prior to the sample. Changes in transmittance were monitored using a CMOS spectrometer (Avantes StarLine AvaSpec-ULS2048CL-EVO) within the 500–700 nm wavelength range. This spectral window was selected to capture the dominant variations in the visible region during illumination. The schematic of the experimental setup is shown in Fig. S1 in the Supporting Information. Wavelength-dependent photochromic changes were evaluated by measuring UV–Vis–NIR transmittance spectra (320–2500 nm) before and after 10 min of illumination with a 365 nm flashlight.

## 3. Results and discussion

### *3.1. Structure of YHO films*

#### *3.1.1. XRD out-of-plane orientation analysis*

Out-of-plane HR-XRD patterns measured in quasi-symmetric $\omega/2\theta$ scan mode over the $2\theta$ range of 20°–75° (with an $\omega$ offset of 0.1° or 0.5° to minimise the contribution from the intense substrate peaks) for the optimised YHO films (samples S-YSZ111 and S-YSZ100-1), exhibiting the best lattice matching with the substrates, deposited on YSZ(111) and YSZ(100), are presented in Fig. 1a and 1b, respectively. Two broad diffraction peaks, located to the left of the narrow substrate peaks, are attributed to the YHO films. Assuming that the film adopts the same crystal system as the substrate (cubic), and based on the previously reported indexing of YHO films deposited on glass [29] and Si substrates [8], these peaks were assigned to the (111) and (222) lattice planes of films on YSZ(111), and to the (200) and (400) lattice planes of films on YSZ(100).

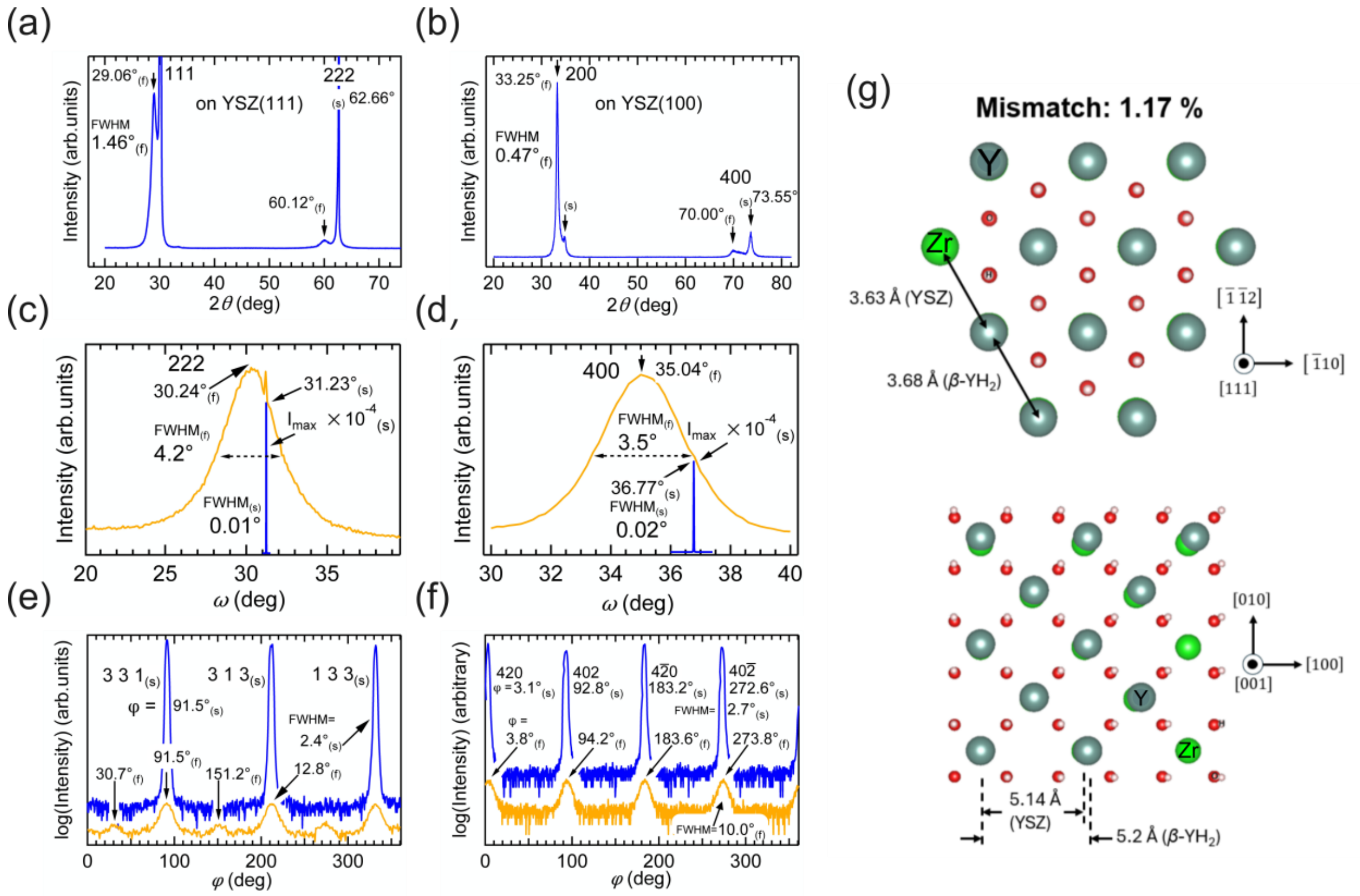

**Fig. 1.** Selected area HR-XRD patterns of YHO films deposited on YSZ substrates at 200 °C (samples S-YSZ111 and S-YSZ100-1). (a,b) Quasi-symmetric $\omega/2\theta$ scans of YHO films deposited on (a) YSZ(111) and (b) YSZ(100) substrates. (c,d) $\omega$ (rocking curve) scans of the 222 and 400 peaks for films on (c) YSZ(111) and (d) YSZ(100) substrates (the narrow peaks

correspond to substrate peaks, illustrating their position and instrumental broadening). (e,f) asymmetric coplanar $\varphi$-scans of the 311 and 420 peaks for films deposited on (e) YSZ(111) and (f) YSZ(100) substrates, respectively. In all panels, labels (f) and (s) denote film and substrate peaks, respectively. Peak and goniometer axis positions ($2\theta$, $\varphi$, or $\omega$, in degrees) are indicated above the peaks or by arrows, and the full width at half maximum of peaks is denoted as FWHM. (g) Schematic illustration of the epitaxial relationship between the $\beta$-$YH_2$ precursor film and the YSZ substrate for the (111) and (001) orientations, showing the corresponding lattice mismatch. Y, Zr, H, and O atoms are shown as grey, green, light pink, and red spheres, respectively.

The absence of additional film-related diffraction peaks suggests that the films were preferentially oriented, with their lattices aligned out-of-plane relative to the substrate, consistent with the epitaxial relationships illustrated schematically in Fig. 1g. The corresponding out-of-plane lattice parameters were determined to be $a$ = 0.5320(5) nm and $a$ = 0.5380(5) nm for films deposited on YSZ(111) and YSZ(100), respectively. These values are consistent with the lattice parameter of transparent yttrium hydride ($a$ = 0.5346 nm) reported by T. Mongstad et al. [29], as well as with predicted lattice constants for stoichiometric YHO with $P\bar{4}3m$ symmetry (0.5361–0.5385 nm) [30].

The precursor $\beta$-$YH_x$ phase undergoes post-deposition oxidation, leading to the formation of photochromic YHO while largely preserving the *fcc* yttrium sublattice. However, Y K-edge EXAFS measurements have revealed a splitting of the second Y coordination shell into two distinct Y–Y distances, which is inconsistent with ideal cubic symmetry and is therefore attributed to local structural relaxation around anion vacancies [4]. During oxidation, oxygen atoms preferentially occupy tetrahedral interstitial sites, as supported by a combined X-ray and neutron diffraction study [10] as well as Madelung energy calculations [31]. Consequently, hydrogen is displaced either to less favourable octahedral sites or partially released from the lattice. The resulting anion distribution and degree of ordering are therefore strongly dependent on both the O/H atomic ratio and the size of the rare-earth cation, as observed in related oxyhydride systems [11,32]. Notably, the O/H ratio can vary over a relatively wide range and is highly sensitive to the deposition conditions, most critically to the working pressure during the deposition of the reactively sputtered $\beta$-$YH_x$ precursor.

Both theoretical and experimental studies also indicate that hydrogen-rich yttrium oxyhydride compositions (O/Y < 1, H/Y > 1) exhibit a predominantly disordered anion sublattice [10,11]. Although DFT predicts that stoichiometric YHO may adopt energetically favourable anion-ordered structures such as $F\bar{4}3m$ or $P\bar{4}3m$ [30], these have not been conclusively verified experimentally. This suggests that any anion-ordering is likely limited to short-range correlations rather than long-range periodicity. Furthermore, although XRD provides structural information on the yttrium and oxygen sublattices, its very low sensitivity to hydrogen prevents a conclusive determination of the anion arrangement. Therefore, even though the measured lattice parameters are consistent with the $P\bar{4}3m$ space group, its presence cannot be unambiguously confirmed.

Although the lattice parameters of films deposited on differently oriented substrates do not differ significantly, the diffraction peak broadening varies by approximately a factor of three. Assuming that the broadening is dominated by finite coherent domain size, crystallite sizes estimated using the Scherrer equation [33] are 7.8(4) nm and 20(1) nm for films deposited on YSZ(111) and YSZ(100), respectively. Since microstrain may also contribute to the observed broadening, these values should be regarded as first-order estimates.

The quality of epitaxy is commonly assessed by the mosaicity of the film, which can be characterised by the full width at half maximum (FWHM) of diffraction peaks measured in $\omega$ (rocking curve) scans. The mosaicity of the films was found to be 4.2° and 3.5° for samples deposited on YSZ(111) and YSZ(100) substrates, respectively (Fig. 1c and 1d). Although substantially larger than the ≈0.1–0.2° values typically observed for high-quality semiconductor epitaxy, these results confirm a well-defined out-of-plane orientation and indicate epitaxial-like growth characterised by a pronounced mosaic spread [34].

The out-of-plane XRD patterns using Bragg-Brentano optics of the samples are presented in Fig. S2 (Supporting Information). Compared with the $\beta$-$YH_2$ reference, XRD peaks of the YHO films are consistently shifted towards lower diffraction angles, reflecting an increase in the lattice parameter associated with oxygen incorporation, in line with earlier studies [29,35,36]. This lattice expansion is commonly attributed not only to anion substitution but also to the lower polarizability of $O^{2-}$ compared to $H^-$, which modifies the local bonding environment and induces lattice strain despite their similar ionic radii [31].

The importance of optimising the deposition conditions to achieve lattice matching with the YSZ substrates is illustrated by comparing Fig. S2b and S2c. For YSZ(100), promoting lattice matching required a lower working pressure of 0.93 Pa, reduced discharge power of 100 W, and a higher $H_2$/Ar ratio of 0.05 (sample S-YSZ100-1), compared to 1.65 Pa, 200 W, and an $H_2$/Ar ratio of 0.02 (sample S-YSZ100-2). Under the latter conditions, additional reflections corresponding to (111) and (222) lattice planes were observed alongside the 200 and 400 peaks, indicating a less well-defined out-of-plane orientation.

The films deposited on amorphous $SiO_2$ and soda-lime glass substrates are polycrystalline and exhibit varying degrees of preferred orientation. Sample S-SiO-200C is close to a randomly oriented polycrystalline structure ($TC_{111}$ = 1.1), whereas the RT deposited films exhibit more pronounced texture. Sample S-SL-RT shows a preferred 111 orientation ($TC_{111}$ = 1.7), while sample S-SiO-RT exhibits a dominant 100 texture ($TC_{200}$ = 2.5). The main difference between these two samples is the $H_2$/Ar gas flow ratio during $\beta$-$YH_x$ deposition, which influences adatom surface mobility on different lattice planes and grain-boundary kinetics through hydrogen adsorption, incorporation, and reaction processes, thereby modifying competitive grain growth and the resulting preferred crystallographic orientation of the YHO films. The crystallite sizes, estimated using the Scherrer equation, are 11, 13, and 13 nm for samples S-SiO-200C, S-SiO-RT, and S-SL-RT, respectively. These results, together with the comparison to the YSZ samples, indicate that neither the elevated deposition temperature of 200 °C nor the crystallinity of the substrate leads to a significant increase in crystallite size during $\beta$-$YH_x$ deposition.

### *3.1.2. XRD in-plane lattice orientation analysis*

Asymmetric coplanar $\varphi$-scan diffraction analysis (rotation of the sample about the surface normal when incident and exit angles are not equal) was employed to verify the assumed lattice symmetry of the film material and to determine the mutual in-plane orientation of the film and substrate lattices. For the film deposited on YSZ(111), the {331} family of planes, corresponding to peaks at $2\theta$ = 78.70°, was selected for $\varphi$-scan analysis, as this is the first accessible family that shows small enough angle with (111) plane for allowing recording of diffraction reflections in asymmetric coplanar diffraction geometry. Similarly, for the film deposited on YSZ(100), the {210} family of planes (observed via the second-order 420 peaks at $2\theta$ = 84.03°) was used.

All observable film peaks (three for YSZ(111) and four for YSZ(100)) were found to be angularly aligned with the corresponding substrate peaks (Fig. 1e and 1f). Although the film peaks exhibited a broadening (~10–13°) that was approximately 3–5 times greater than the instrumental broadening of the $\varphi$-scan peaks (~2–3°), the results indicate a modest degree of in-plane lattice alignment between the films and the substrates. The following in-plane orientational relationships were identified: $[\bar{1}01]$ YHO(111) // $[\bar{1}01]$ YSZ(111) and $[100]$ YHO(100) // $[100]$ YSZ(100), for films deposited on YSZ(111) and YSZ(100), respectively.

For the films on YSZ(111), an additional set of weaker peaks was observed, located midway between the primary peaks. These peaks indicate the presence of a secondary set of crystallites rotated by 60° with respect to the dominant in-plane orientation. This secondary orientation can be rationalised by considering the atomic arrangement in the (111) plane of the YHO structure (space group $Fm\bar{3}m$, lattice parameter $a$ = 0.53226 nm [10]). Although the overall structural symmetry in this plane is threefold, the anion sublattice exhibit an apparent sixfold rotational symmetry when anions occupy both tetrahedral and octahedral sites (Fig. S3 in the Supporting Information), which enables the formation of crystallites rotated by 60° relative to those whose orientation is determined by the *fcc* yttrium sublattice.

The presence of reciprocal lattice points in the equatorial plane for the film on YSZ(100) enabled determination of the in-plane lattice parameter and crystallite size. The in-plane $\theta$–$2\theta$ scan (or $\varphi$–$2\theta\chi$ scan according to the SL goniometer notation) was performed by rotation of sample and detector in 1:2 relation around sample surface normal using X-ray beam incident and exit angles of 0.35 and ≈0.7°, respectively. The position of the 002 peak at $2\theta$ = 33.72° (Fig. 2) corresponds to an in-plane lattice parameter of $a$ = 0.5310(5) nm, indicating a slight in-plane lattice compression compared to the out-of-plane direction ($a$ = 0.538 nm). This behaviour is consistent with the smaller lattice parameter of the template serving YSZ substrate ($a$ = 0.515 nm) compared to the relaxed lattice parameter of yttrium oxyhydride ($a$ ≈ 0.520–0.538 nm) [37].

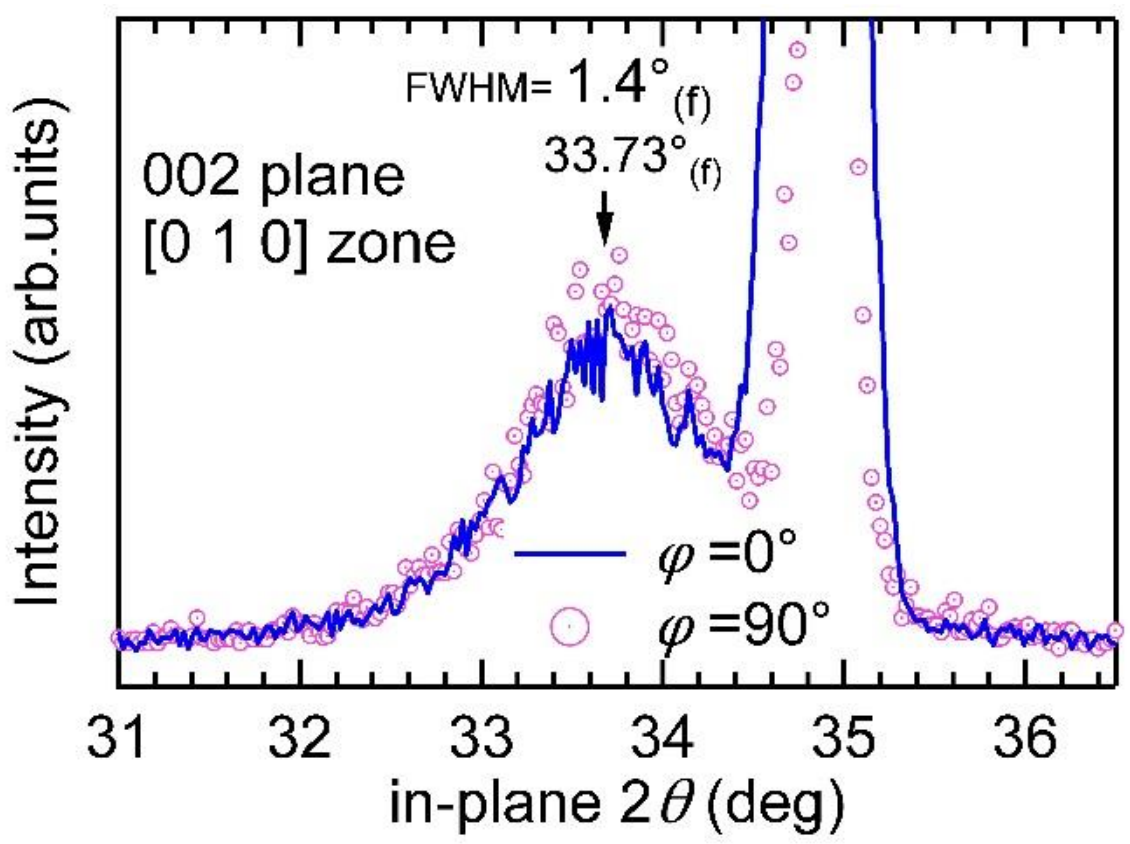

**Fig. 2.** In-plane $\theta$–$2\theta$ scans of film on YSZ(100) in the vicinity of the 002 peak, measured at sample rotation angles corresponding to the [100] and [010] zone axes. Position and FWHM of the film peak are shown at the top of the peak. The instrumental broadening, estimated from FWHM of 002 peak of substrate located at ~35°, was approximately 0.2°.

The in-plane crystallite size, estimated from the broadening of the 002 reflection, was 10(1) nm, again without accounting for possible microstrain contributions. Considering that the crystallite size in the out-of-plane direction is approximately twice as large, the crystallites exhibit an anisotropic morphology, elongated along the surface normal. The characteristic structural parameters obtained from HR-XRD are summarised in Table 2.

**Table 2.** Structural parameters in in-plane (∥) and out-of-plane (⊥) direction of YHO films deposited on YSZ(111) and YSZ(100) substrates, determined from HR-XRD analysis. The films were deposited at 200 °C with working pressures of 1.80 Pa (YSZ(111)) and 0.95 Pa (YSZ(100)), and an $H_2$/Ar flow ratio of 0.05. Unit cell parameters $a$ and X-ray apparent crystallite sizes $\langle D \rangle_v$ were obtained from $2\theta$ scans, while mosaicity values correspond to the FWHM of $\omega$ rocking curves. Dashes (–) indicate parameters that could not be determined.

| Sample | Substrate | $a$ ⊥ (nm) | $a$ ∥ (nm) | $\langle D \rangle_v$ ⊥ (nm) | $\langle D \rangle_v$ ∥ (nm) | Mosaicity ⊥ (°) |
|---|---|---|---|---|---|---|
| S-YSZ111 | YSZ(111) | 0.5320(5) | – | 7.8(4) | – | 4.2 |
| S-YSZ100-1 | YSZ(100) | 0.5380(5) | 0.5310(5) | 20(1) | 10(1) | 3.5 |

### 3.1.3. SEM of YHO films

Top-view SEM images of the YHO films are presented in Fig. 3, together with cross-sectional images of the films deposited on the YSZ substrates. The film deposited on YSZ(111) exhibits a non-uniform surface morphology with a loosely packed columnar grain structure separated by pronounced intercolumnar voids (Fig. 3a), indicative of limited adatom surface mobility and strong shadowing during growth. The surface morphology is characterised by grains with faceted surfaces, where triangular facets are frequently observed at the topmost surface. Although the (111) surface of the cubic $Fm\bar{3}m$ structure possesses threefold symmetry, a hexagonal morphology would generally be expected under equilibrium growth conditions. The observed triangular facets therefore suggest kinetically limited growth, potentially associated with corner-crossing asymmetry and anisotropic adatom diffusion along crystallographically distinct step edges [38,39]. A similar mechanism has been proposed for metallic *fcc*(111) systems, where kinetic growth anisotropies promote the formation of triangular islands instead of the equilibrium hexagonal morphology. However, the microscopic origin of the anisotropy in the $YH_2$/YHO system remains unclear and may be influenced by the presence of hydrogen in the lattice. In addition, distinct star-like features are observed, highlighted by dashed red circles in Fig. 3a and clearly visible at higher magnification. These features arise from crystallites rotated by 60° with respect to each other, consistent with the secondary in-plane orientation identified in the $\varphi$-scan analysis (Fig. 1e). The formation of such rotated domains can be attributed to the disordered anion sublattice, which allows energetically similar growth variants.

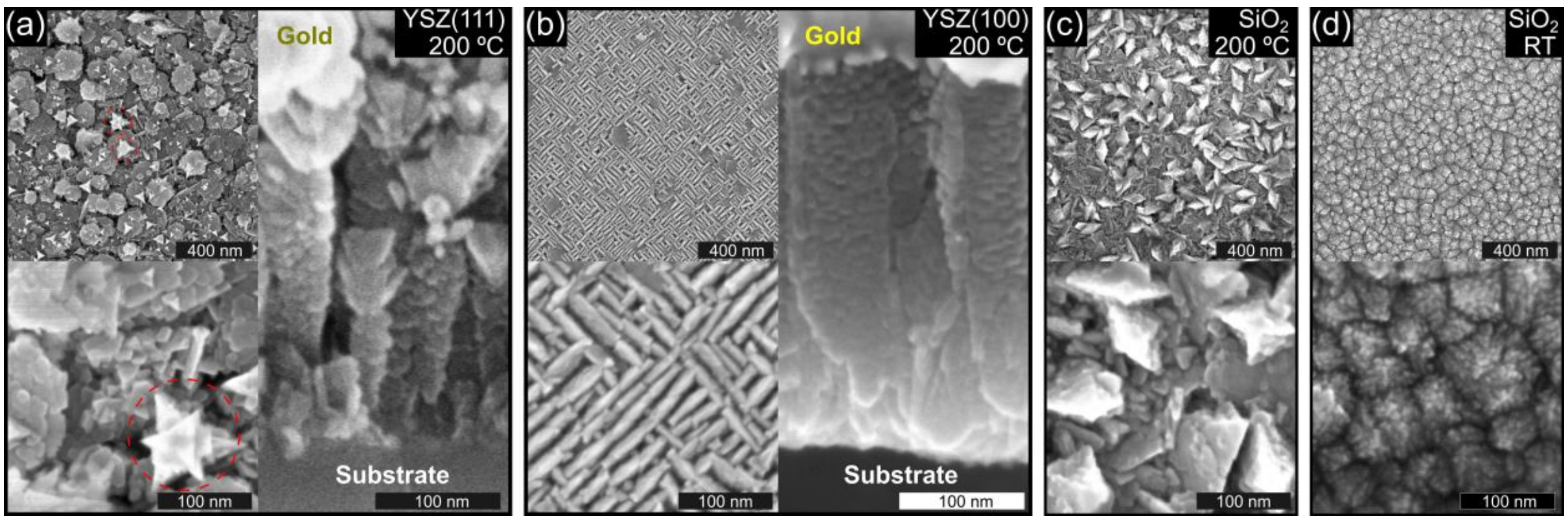

**Fig. 3.** Top-view SEM images at two magnifications of YHO films deposited on (a) YSZ(111) (sample S-YSZ111), (b) YSZ(100) (sample S-YSZ100-1), and (c) amorphous $SiO_2$ (sample S-SiO-200C) at a substrate temperature of 200 °C, and on (d) amorphous $SiO_2$ at room temperature (sample S-SiO-RT). The dashed red circles in (a) highlight representative star-like

surface features. Cross-sectional SEM images of the films deposited on YSZ(111) and YSZ(100) are also included in (a) and (b), respectively.

In contrast, the YHO film deposited on YSZ(100) exhibits a markedly different microstructure (Fig. 3b). The surface is composed of elongated, rod-like crystallites arranged along two orthogonal directions, forming a highly ordered cross-hatched texture. While the two orthogonal growth directions are consistent with the fourfold symmetry of the YSZ(100) substrate, the elongated morphology is not expected under equilibrium growth conditions, which would favour square-shaped surface features. Instead, the rod-like crystallites suggest anisotropic growth kinetics, potentially arising from corner-crossing asymmetry or anisotropic edge diffusion specific to the $YH_2$/YHO–YSZ(100) system. Despite the larger crystallite size and lower mosaicity compared with the film deposited on YSZ(111), the film appears porous based on qualitative observations, with voids between adjacent crystallites. In addition to the dominant texture, irregularly oriented features are observed, indicating the presence of misoriented grains. Finally, both YSZ(111) and YSZ(100) surfaces are known to undergo surface reconstruction and exhibit yttrium segregation as well as defect-rich surface structures under different preparation conditions [40–42]. These surface characteristics may locally modify adatom diffusion pathways and thereby contribute to the observed growth anisotropy.

The YHO films deposited on amorphous $SiO_2$ (Fig. 3c and 3d) exhibit seemingly denser and more isotropic surface morphologies. For the film grown at 200 °C (Fig. 3c), the surface consists of faceted grains with pyramidal morphology and well-defined edges. In contrast, the film deposited at RT (Fig. 3d) exhibits a compact, uniformly fine-grained surface morphology with rounded grains. The surface appears smoother, and the grains themselves display a nanocrystalline substructure. Unlike the films deposited on YSZ substrates, neither sample exhibits substrate-induced anisotropic grain alignment or epitaxial-like faceting. Nevertheless, XRD analysis reveals varying degrees of preferred crystallographic orientation (Section 3.1), indicating that texture development on amorphous substrates is governed primarily by growth kinetics.

*3.2. Composition of YHO*

The composition depth profiles, determined by ToF-E ERDA, of the studied samples are shown in Fig. 4, and the corresponding atomic concentrations and stoichiometries are summarised in Table 3. Both oxygen and hydrogen concentrations vary significantly in the near-surface region as a result of oxidation and are strongly influenced by the film microstructure and the duration of exposure to ambient conditions. To obtain representative values, the compositions were evaluated from the deeper regions of the films, where the elemental distributions are more homogeneous. An additional source of uncertainty arises from elemental loss during measurement, particularly for hydrogen, which can be significant under the experimental conditions. For the optimised samples deposited close to $P_c$, the O/Y and O/H atomic ratios range from 0.71 to 0.82 and 0.33 to 0.44, respectively. In addition, carbon, fluorine, and nitrogen impurities were detected with a total concentration ranging from 2.7 to 4.0 at.%, with their relative proportions varying among the samples (see Table S1 in the Supporting Information), although the origin of these variations remains unclear. Comparable concentrations of the same impurities have previously been reported for YHO and other rare-earth oxyhydride (REHO) thin films deposited using different sputtering systems [2,43], although their potential influence on the photochromic properties has not been clarified. The fluorine contamination may originate from the Y target [44]. For clarity, these impurity contributions were excluded from the depth profiles shown in Fig. 4.

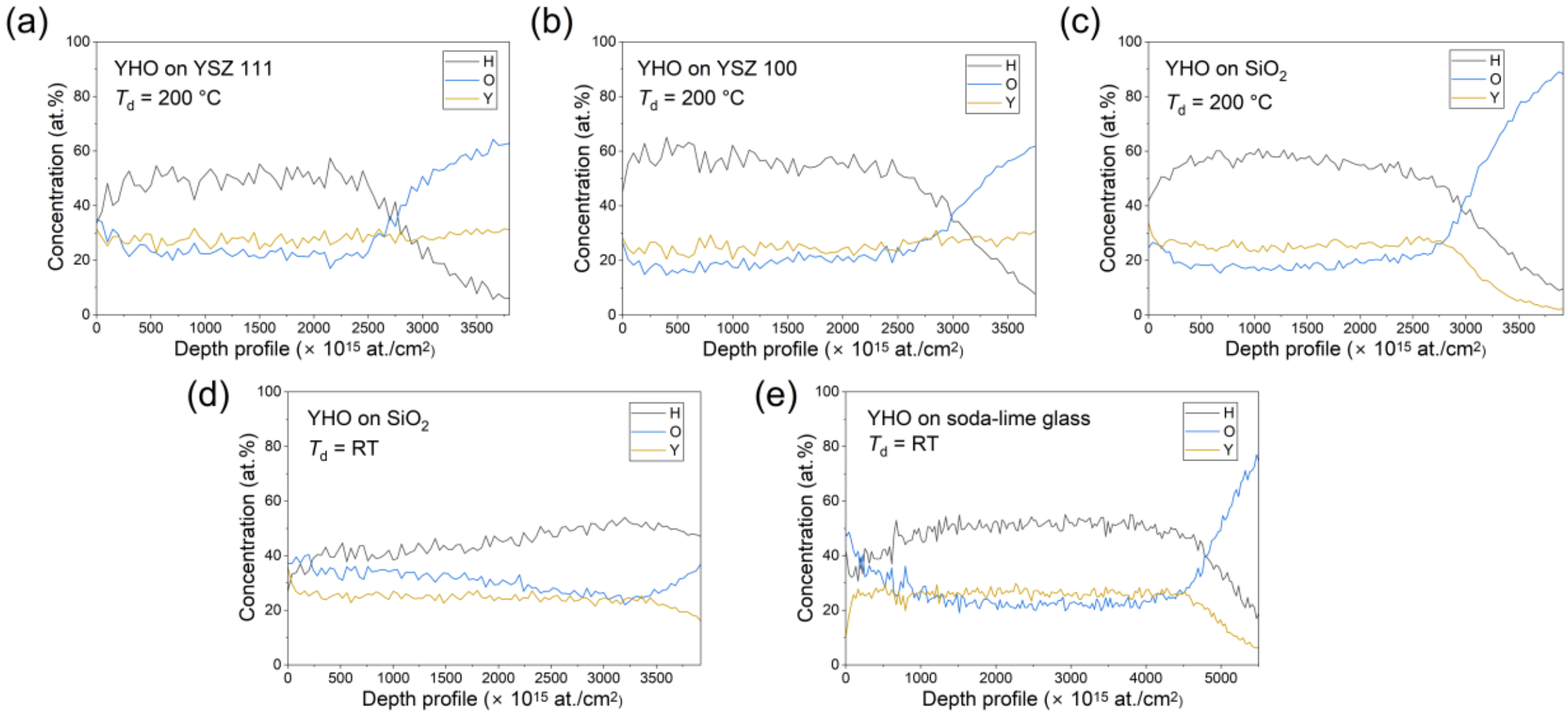

**Fig. 4.** Composition depth profiles of YHO thin films deposited on YSZ, $SiO_2$, and soda-lime glass substrates at deposition temperatures of 200 °C and RT, determined by ToF-E ERDA. The atomic concentrations of hydrogen (H), oxygen (O), and yttrium (Y) are shown as a function of areal density for samples (a) S-YSZ111, (b) S-YSZ100-1, (c) S-SiO-200C, (d) S-SiO-RT, and (e) S-SL-RT. In the case of films deposited on YSZ substrates (a,b), the Y signal does not decrease in the substrate region due to overlap with the Zr signal arising from the similar atomic masses of Y and Zr.

**Table 3.** Composition (at.%), derived stoichiometry, O/H atomic ratio, yttrium atomic density ($n_Y$), and mass density ($\rho$) of YHO films determined by ToF-E ERDA, with $n_Y$ and $\rho$ derived in combination with film thickness measurements by spectroscopic ellipsometry.

| Sample | Composition (at.%) | | | Stoichiometry | | | O/H | $n_Y \times 10^{22}$ (at cm$^{-3}$) | $\rho$ (g cm$^{-3}$) |
|---|---|---|---|---|---|---|---|---|---|
| | Y | H | O | Y | H | O | | | |
| S-YSZ111* | 27.2 | 50.5 | 22.4 | 1.00 | 1.85 | 0.82 | 0.44 | 2.4 | 4.1 |
| S-YSZ100-1* | 24.0 | 57.1 | 18.9 | 1.00 | 2.38 | 0.79 | 0.33 | 2.4 | 4.2 |
| S-SiO-200C* | 25.7 | 55.9 | 18.3 | 1.00 | 2.17 | 0.71 | 0.33 | 2.2 | 3.8 |
| S-SiO-RT | 23.5 | 47.0 | 29.5 | 1.00 | 2.00 | 1.26 | 0.63 | 2.2 | 4.0 |
| S-SL-RT* | 23.3 | 55.2 | 21.5 | 1.00 | 2.23 | 0.82 | 0.37 | 2.4 | 4.1 |

* Optimised samples

Two compositional formulas, $YH_{2-\delta}O_\delta$ [2,43] and $YH_{3-2x}O_x$ [45], have been used to describe the film composition in relation to the O/H ratio. The $YH_{2-\delta}O_\delta$ formula reflects the gradual oxidation, and thus the multiphase formation, of an initially formed yttrium dihydride ($\beta$-$YH_{1.8-2.3}$) precursor, with the yttrium oxidation state evolving from +2 towards +3 in order to maintain charge neutrality in the presence of $H^-$ and $O^{2-}$ anions. In the transparent state of photochromic $YH_{0.7}O_{2.2}$, yttrium has been reported to exhibit an effective oxidation state of approximately +2.5 [46]. Within this framework, and considering that octahedral sites are energetically less favourable, anions are expected to predominantly occupy tetrahedral interstitial sites. In contrast, the $YH_{3-2x}O_x$ formula assumes yttrium in a +3 oxidation state and requires partial occupation of octahedral sites by hydrogen at low oxygen contents ($x < 1$). However, based on bulk thermodynamic considerations, the formation of $\gamma$-$YH_3$ during magnetron sputtering is generally not expected, as the hydrogen partial pressure under typical deposition conditions is well below the equilibrium pressure required for the $\beta$-$YH_2$ to $\gamma$-$YH_3$ phase transition [47,48].

The obtained compositions (Table 3) do not conform to either of the proposed compositional formulas, with the measured hydrogen contents consistently exceeding those predicted by both formulas. This suggests that, during oxidation, hydrogen is initially displaced into octahedral sites rather than being immediately released from the lattice. In addition, charge neutrality may be maintained through the formation of OH groups [16] or water in porous films, which cannot be distinguished by ERDA. Furthermore, beyond the near-surface region, the composition is not completely uniform throughout the film thickness but exhibits a compositional gradient that depends on the microstructure and, consequently, on the oxidation process, which may be influenced by the deposition conditions and the type of substrate.

The total areal atomic density (at $cm^{-2}$) obtained from ERDA was converted to volumetric atomic density (at $cm^{-3}$) using the independently measured film thickness. The film mass density ($\rho$) and yttrium atomic density ($n_Y$) were subsequently derived from the volumetric atomic density and elemental composition (Table 3). These values are subject to relatively large uncertainties arising from the absence of a sharp film-substrate interface in the ERDA depth profiles, possible variations in film thickness between the ERDA and thickness measurement positions, and uncertainties associated with the thickness and compositional analyses. $n_Y$ can serve as an indicator of the density of the as-deposited $\beta$-$YH_x$ precursor, since the Y content is expected to remain unchanged during post-deposition oxidation, whereas hydrogen may be released from the film. Interestingly, although the SEM images (Fig. 3) suggest that the epitaxial-like YHO films are more porous than those deposited on amorphous substrates, the calculated $n_Y$ values do not support this qualitative observation. Among the optimised films, $n_Y$ is approximately $2.4 \times 10^{22}$ $cm^{-3}$, except for S-SiO-200C, which exhibits a lower value of $2.2 \times 10^{22}$ $cm^{-3}$. The corresponding mass densities range from 3.8 to 4.1 g $cm^{-3}$. In comparison, calculated densities reported for various YHO phases in Ref. [30] range from 4.33 to 4.75 g $cm^{-3}$. Taking these calculated values as the fully dense reference suggests an approximate film porosity of 3–20 %. Since post-deposition oxidation is uncontrolled and only partly self-limiting [49], the density and microstructure of the $\beta$-$YH_x$ precursor strongly influence oxygen incorporation and, consequently, the final YHO composition. However, the present density estimates do not establish a direct correlation between film density and oxygen content, which may partly reflect their relatively large uncertainties. Increased deposition pressure has previously been shown to promote oxygen incorporation, with excessive oxidation occurring at pressures above the narrow range providing optimal photochromic performance [7].

*3.3. Optical and photochromic properties of YHO*

All YHO thin films are transparent from the optical absorption edge at approximately 450 nm up to 2500 nm, as shown by the transmittance spectra in Fig. 5a, consistent with the optical behaviour of photochromic YHO films. The average transmittance in the visible spectral range ($T_{400–750}$), together with the corresponding solar ($T_{sol}$) and luminous ($T_{lum}$) transmittance values calculated according to equation $T_{sol/lum} = \frac{\int T(\lambda)\varphi_{sol/lum}(\lambda)d\lambda}{\int \varphi_{sol/lum}(\lambda)d\lambda}$ are summarised in Table 4. The transmittance of the YHO films, determined after correcting for the transmittance of the corresponding substrates, varies between approximately 57 and 82 %, depending on the sample. Within the investigated sample set, the optical transparency correlates well with the oxygen content of the films (Tables 3 and 4).

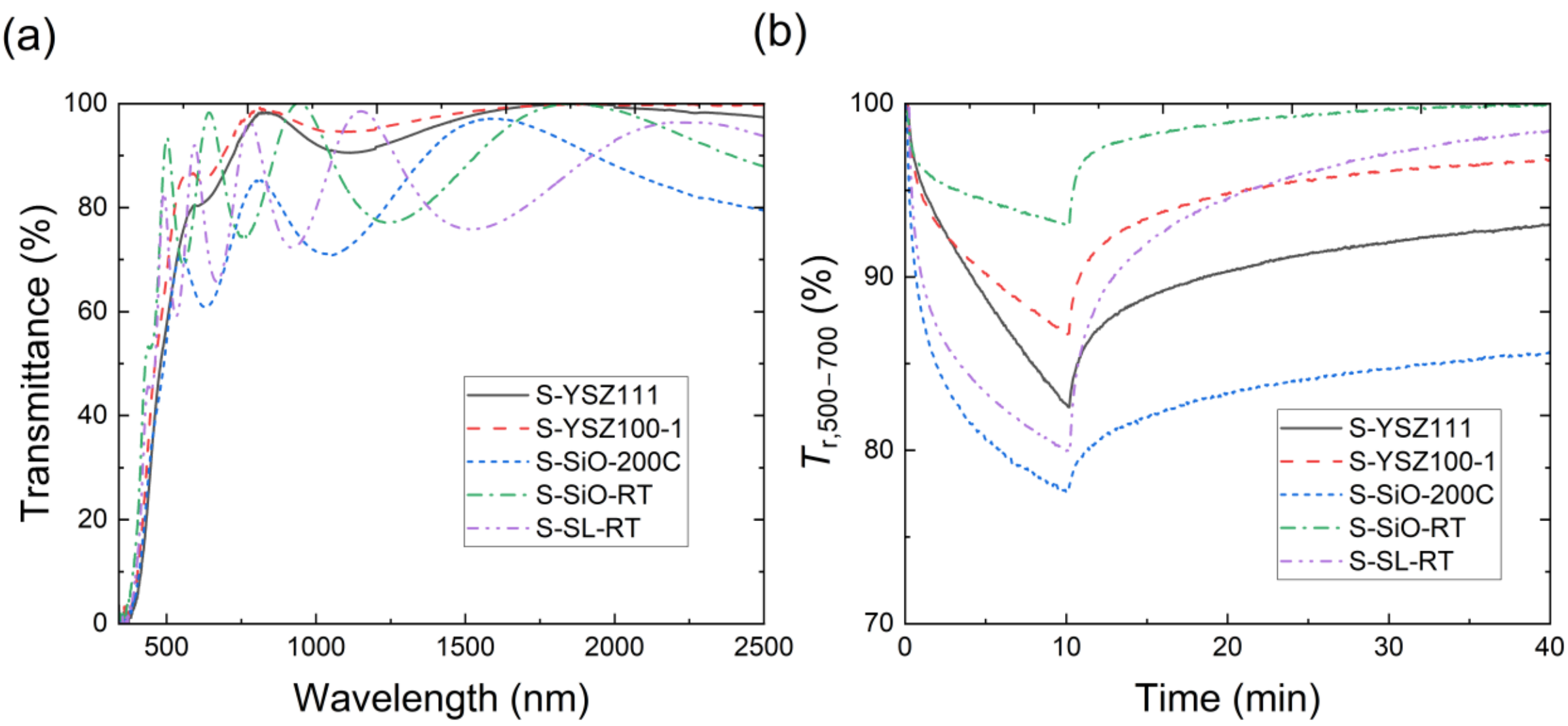

**Fig. 5.** (a) Optical transmittance spectra of YHO films deposited on YSZ, $SiO_2$, and soda-lime glass substrates at RT and 200 °C after subtracting the corresponding substrate contribution. (b) Relative time evolution of the average transmittance in the 500–700 nm wavelength range ($T_{r,500–700}$) during illumination with a solar simulator for films aged for four months after deposition. The legend indicates the corresponding samples defined in Table 1.

The optical band gap of the investigated films, determined from Tauc plots assuming direct allowed transitions, varies between 2.9 and 3.1 eV, consistent with previously reported values for photochromic YHO thin films [7,16]. The corresponding Tauc plots and linear fitting regions are provided in Fig. S4 of the Supporting Information. Within the present sample set, however, no clear correlation is observed between $E_g$ and either the O/Y or O/H atomic ratio, despite composition-dependent variations in $E_g$ reported previously [45]. This apparent discrepancy may partly arise from uncertainties in both the compositional analysis and the determination of $E_g$, with the uncertainty in the latter estimated to be approximately 0.05–0.10 eV. Moreover, crystallite-size effects have previously been discussed in relation to the optical band gap of YHO [3]. Consequently, the relatively modest variations in $E_g$ observed here cannot be unambiguously attributed to a single compositional or structural parameter.

**Table 4.** Optical and photochromic properties of YHO thin films deposited on YSZ, $SiO_2$, and soda-lime glass substrates at RT and 200 °C. The table summarises the average transmittance in the visible spectral range ($T_{400–750}$), solar transmittance ($T_{sol}$), luminous transmittance ($T_{lum}$), optical band gap ($E_g$), relative photochromic contrast in the 500–700 nm wavelength range ($\Delta T_r$), $\Delta T_r$ normalised to a film thickness of 325 nm ($\Delta T_{r,n}$), and effective bleaching time constant ($\tau_{b,eff}$).

| Sample | $T_{400–750}$ (%) | $T_{sol}$ (%) | $T_{lum}$ (%) | $E_g$ (eV) | $\Delta T_r$ (%) | $\Delta T_{r,n}$ (%) | $\tau_{b,eff}$ (min) |
|---|---|---|---|---|---|---|---|
| S-YSZ111* | 66 | 77 | 73 | 2.9 | 18 | 17 | 70 |
| S-YSZ100-1* | 73 | 81 | 82 | 3.0 | 14 | 14 | 11 |
| S-SiO-200C* | 57 | 66 | 65 | 3.0 | 22 | 20 | 366 |
| S-SiO-RT | 76 | 78 | 78 | 3.1 | 7 | 6 | 4 |
| S-SL-RT* | 67 | 73 | 74 | 3.1 | 20 | 15 | 8 |

* Optimised samples

All investigated films exhibited photochromic behaviour, confirming the formation of the photochromic YHO phase in all cases. The relative photochromic response ($\Delta T_r = \Delta T / T_0$, where $T_0$ is the transmittance before illumination) of the YHO thin films was evaluated by illuminating the samples from the film surface side with a solar simulator for 10 min (radiant exposure of 60 J cm$^{-2}$) while monitoring the evolution of the average optical transmittance in the 500–700 nm wavelength range (Fig. 5b). Since the second step of YHO synthesis involves spontaneous post-deposition oxidation in air until a metastable state is reached, the photodarkening measurements were performed twice: immediately after deposition and after an ageing period of four months. The latter results are presented in Fig. 5b and Table 4. A decrease in the photochromic contrast was observed after ageing for all samples, although to different extents, which is attributed to continued oxidation and gradual hydrogen loss from the films [49,50]. Wavelength-dependent transmittance spectra (320–2500 nm) measured in the clear state and after 10 min of illumination with a 365 nm flashlight are provided in Fig. S5 of the Supporting Information.

Since $\Delta T_r$ exhibits an approximately linear dependence on film thickness as reported in Ref. [3] within the range of 325–550 nm studied here, the measured contrasts cannot be directly compared without accounting for differences in film thickness. Therefore, the $\Delta T_r$ values were normalised to a reference thickness of 325 nm using the thickness dependence reported in Ref. [3], yielding the normalised relative photochromic response ($\Delta T_{r,n}$) presented in Table 4. However, these normalised values are subject to considerable uncertainty because illumination conditions and other sample-specific properties may influence the dependence of the photochromic response on film thickness.

The $\Delta T_{r,n}$ of the optimised samples ranges from 14 to 20 %. The highest $\Delta T_{r,n}$ of 20 % is observed for S-SiO-200C, which also exhibits the lowest O/Y atomic ratio of 0.71 (Table 3), in agreement with previously reported relationships between YHO composition and photochromic activity [51]. Among the remaining optimised samples, S-YSZ111, S-YSZ100-1, and S-SL-RT, which exhibit similar O/Y atomic ratios, S-YSZ111 shows the highest $\Delta T_{r,n}$ of 17%. This sample also exhibits a slightly lower $E_g$ of 2.9 eV, which enhances absorption of the illumination spectrum in the violet region and may thereby contribute to its somewhat higher photochromic response, as suggested in Ref. [5]. Nevertheless, given the relatively small differences in both $\Delta T_{r,n}$ and $E_g$, this contribution cannot be unambiguously separated from other compositional and microstructural effects. Overall, despite the pronounced differences in texture and morphology among the optimised films deposited on different substrates and at different temperatures, no correspondingly large variation in the photochromic contrast is observed. This observation is consistent with previous reports on photochromic YHO films ($\Delta T > 10$ %), where the relative intensity ratios of the (100) and (200) XRD peaks vary considerably both between different studies [37,52,53] and within individual reports [54,55], without any clear correlation between crystallographic texture and photochromic contrast. This suggests that the film composition, which is strongly influenced by the precursor film density and indirectly controlled by the deposition pressure, remains the primary parameter governing the photochromic contrast.

The effective bleaching time constants ($\tau_{b,eff}$) were determined as described in Ref. [56]. The films deposited at 200 °C exhibit higher $\tau_{b,eff}$ values, corresponding to slower bleaching kinetics, in the range of 11–366 min compared to the RT-deposited films, which exhibit values of 4–8 min. The dependence of $\tau_{b,eff}$ on the photochromic contrast reached during illumination is demonstrated for S-SL-RT in Fig. 6 (blue triangles), where different illumination times, ranging from 5 s to 30 min, were used to achieve different photochromic contrasts, followed by determination of the corresponding $\tau_{b,eff}$ values. A clear, exponential increase in $\tau_{b,eff}$ with increasing contrast is observed for the sample. When $\tau_{b,eff}$ values of the samples from Table 4 are plotted against the corresponding $\Delta T_{r,n}$ in Fig. 6 (red circles), the data are generally consistent with this dependence, although some variation is observed, particularly at higher contrast values. The variation suggests that film-specific properties influence the bleaching kinetics. Overall, these results indicate that $\tau_{b,eff}$ depends strongly on the photochromic contrast reached prior to bleaching and should therefore not be considered solely as an intrinsic material constant of YHO. Nevertheless, the bleaching behaviour can also be influenced by deposition parameters, such as the $H_2$/Ar flow ratio, as demonstrated in Ref. [23].

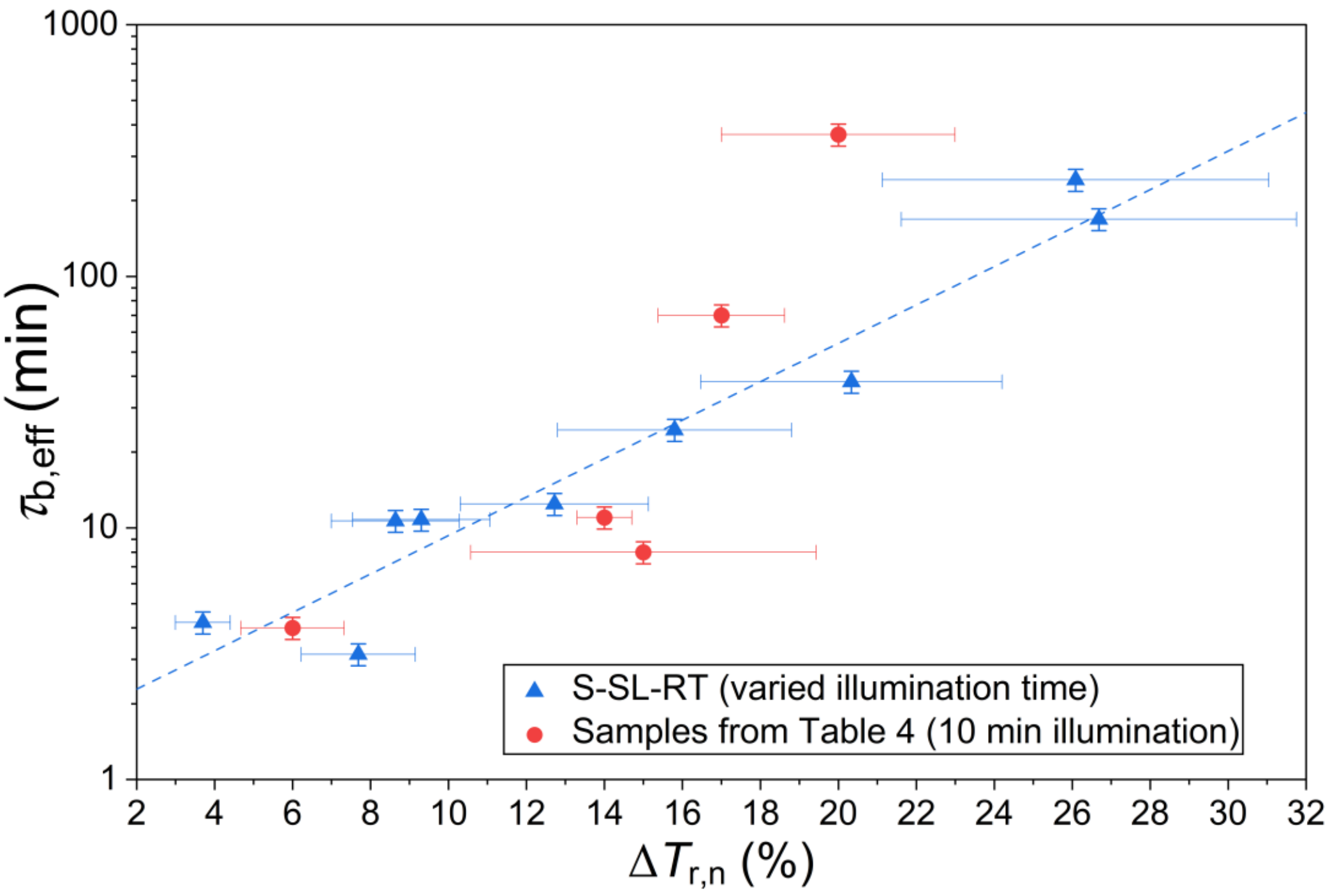

**Fig. 6.** Effective bleaching time constant ($\tau_{b,eff}$) as a function of the normalised relative photochromic contrast ($\Delta T_{r,n}$). Blue triangles represent data for sample S-SL-RT measured after different illumination times, ranging from 5 s to 30 min, resulting in varying photochromic

contrasts, while red circles represent the samples listed in Table 4 after 10 min of illumination with a solar simulator.

## Conclusions

Epitaxial-like YHO thin films were successfully produced on single-crystalline YSZ(111) and YSZ(100) substrates by reactive magnetron sputtering at an elevated substrate temperature of 200 °C, followed by post-deposition oxidation in air. XRD analysis revealed pronounced substrate-orientation-dependent differences in crystallographic quality and lattice anisotropy. The films exhibited mosaicities of 3.5–4.2° and characteristic crystallite sizes of approximately 8–20 nm, with the film on YSZ(100) additionally exhibiting anisotropic crystallite dimensions and residual in-plane compressive strain. The relatively large mosaicity, together with the presence of differently oriented crystallite domains, positions these films in the regime of epitaxial-like growth, whereas the reference films deposited on amorphous substrates exhibited textured polycrystalline structures. Furthermore, $P_c$ increased at the elevated deposition temperature and differed between the two YSZ substrate orientations.

All investigated films exhibited photochromic behaviour; however, despite the pronounced differences in crystallographic texture and morphology, no systematic enhancement of the photochromic contrast was observed for the epitaxial-like films. The optimised films exhibited comparable Y atomic densities despite their distinctly different surface morphologies, leading to similar YHO compositions. Within the investigated sample set, the photochromic contrast is therefore governed primarily by film composition, with no clear independent effect of long-range crystallographic ordering being identified. Furthermore, the effective bleaching time constant increases strongly with the photochromic contrast reached during illumination and should therefore not be regarded as an intrinsic material constant of YHO. Overall, the substrate and deposition temperature strongly affect film growth and microstructure, but their influence on the photochromic response is coupled to the resulting precursor density, oxidation behaviour, and composition. Consequently, optimisation of the deposition conditions is required for each substrate and deposition temperature.

## Declaration of competing interest

The authors declare that they have no known competing financial interests or personal relationships that could have appeared to influence the work reported in this paper.

## Acknowledgements

Financial support was provided by Latvian Council of Science Project No. lzp-2022/2-0454. ToF-E ERDA measurements by MZ were supported from post-doctoral research project 1.1.1.9/LZP/1/24/011. The research team acknowledges the SWEB project 101087367 funded by the HORIZON-WIDERA-2022-TALENTS-01, which provided support for competence development and training. The authors thank A. Ashraf for her assistance with the ToF-E ERDA measurements. HM acknowledges financial support by Estonian Research Council (Grant PRG 2733). KS acknowledges financial support by the Swedish Research Council (Grant No. VR-2021-04113) and the Åforsk foundation (Grant No. 22-150). KS and KM acknowledge financial support by the Jane and Aatos Erkko foundation (Grant No 250022).

During the preparation of this work the author(s) used ChatGPT in order to enhance language clarity and grammatical accuracy. After using this tool/service, the author(s) reviewed and edited the content as needed and take(s) full responsibility for the content of the published article.

**Data availability**

Data will be made available on request.